# A Free Industry-grade Education Tool for Bulk Power System Reliability Assessment

Yongli Zhu[1], *Member*, *IEEE*, Chanan Singh[1], *Fellow*, *IEEE*
[1]Electrical and Computer Engineering Department, Texas A&M University, College Station, USA
yongliz@tamu.edu, singh@ece.tamu.edu

*Abstract*— A free industry-grade education tool is developed for bulk-power-system reliability assessment. The software architecture is illustrated using a high-level flowchart. Three main algorithms of this tool, i.e., sequential Monte Carlo simulation, unit preventive maintenance schedule, and optimal-power-flow-based load shedding, are introduced. The input and output formats are described in detail, including the roles of different data cards and results categorization. Finally, an example case study is conducted on a five-area system to demonstrate the effectiveness and efficiency of this tool.

*Index Terms*—reliability assessment, optimal load shedding, sequential Monte Carlo simulation, unit maintenance schedule.

## I. Introduction

Reliability assessment (also called resource adequacy analysis in certain circumstances) plays a critical role in power system planning and operation [1]. For example, a reliability upgrade is suggested for ERCOT's winterization planning [2][3]. It is also useful in system reconfiguration/restoration, where reliability indices are adopted as main or auxiliary metrics in comparing different plans of topology change [4]. Despite such importance, one obstacle in educating students on this subject is the lack of *free* and *efficient* tools to help students gain direct, hands-on experience.

Commercial tools are powerful, and some may provide specialized reliability assessment modules, e.g., *PowerFactory* [5] and *PSS/E* [6]. However, the financial costs of licensing the pro-versions (free educational versions are available, though limitations on system size and advanced functionalities may be imposed) can be unaffordable for students and junior lecturers. Besides, mastering such commercial tools can be challenging for students due to their intricate user interfaces (since they were originally developed as universal packages rather than *reliability assessment-focused* tools).

Meanwhile, free reliability assessment programs can be sporadically found on the internet, albeit certain limitations exist: 1) lacking advanced features or industry-grade functionalities, such as impacts of (generator) unit *preventive* maintenance, impacts of quarterly or daily events, impacts of in-loop optimal load shedding, impacts of multi-state (>2) probabilistic failure models of generator and line, impacts of interarea power transfer contracts; 2) relying on simple nonsequential Monte Carlo methods; 3) computing in a slow speed (e.g., code written in MATLAB script); 4) lacking detailed user-manual and sustainable software development.

Therefore, the authors of this paper developed a free, industry-grade reliability assessment tool [7] called *NARP* (**N**-**A**rea **R**eliability **P**rogram), which offers all functionalities mentioned above (and more). It is easy to learn for both students and lecturers. Section II illustrates its architecture and main algorithms. Section III describes the formats of its input/output files. Section IV presents an example case study for a five-area system with 120 generators. The final section concludes the paper with future work indicated.

## II. Software Architecture and Main Algorithms

### A. Architecture

*NARP* is a sequential Monte Carlo simulation program, originally developed in *Fortran* to study the planning of generation capacity and tie-line capacity for ERCOT. Its variant versions have been used by Sandia Lab [8], EPRI [9], and NYISO since the 1990s. Its current architecture is shown in Fig. 1, where the left part performs data preprocessing and the right part runs event-based simulations. The current version can be used in either command line mode or graphic-user-interface (GUI) mode (Fig. 2).

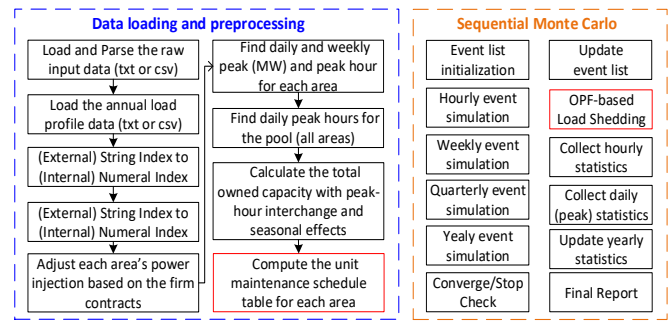

Figure 1. The architecture diagram of the NARP tool

The convergence criteria used in this tool is the "change-of-variance" (denoted as "$\beta$"), as defined by Eq. (1) and (2).

$$F(s_j) \triangleq \text{a mapping function, e.g.} \begin{cases} 0, & s_j \text{ is reliable state} \\ 1, & s_j \text{ is a failure state} \end{cases} \quad (1)$$

$$\beta \triangleq \frac{\sqrt{Var[E[F]]}}{E[F]}, \quad Var[E[F]] = \frac{1}{N} Var[F] \quad (2)$$

where, $s_j$ is the system state sampled at the $j$-th time; $N$ is the total simulation runs; $E[\cdot]$ means the expected value. $F$ is a function mapping the state to a specific reliability index. For example, if Eq. (1) is adopted as $F$, then the expected value in Eq. (2) means the LOLP (Loss of Load Probability).

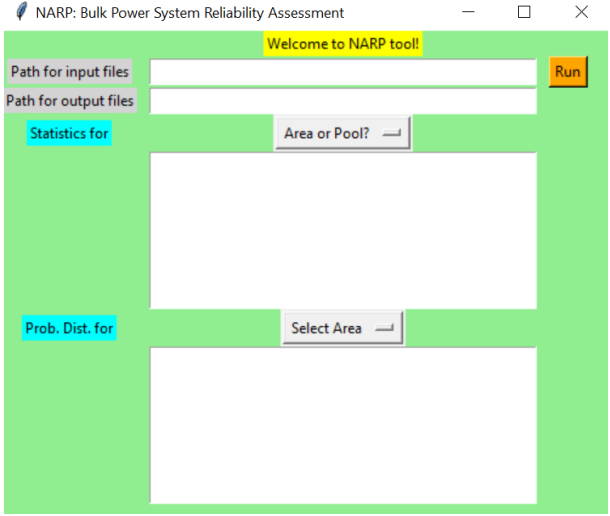

Figure 2. A GUI for the developed *NARP* tool (subject to upgrade)

### B. Main Algorithms

Here, three main algorithms of this tool, i.e., sequential Monte Carlo, unit preventive-maintenance, and OPF (optimal power flow)-based load shedding, are described.

*1) Sequential Monte Carlo Simulation*

Sequential Monte Carlo (SMC) simulation is one genre of stochastic simulation techniques used in reliability assessment [1][10]. Compared to the nonsequential Monte Carlo method, the SMC method is based on event-triggered simulation, e.g., events of quarterly generation adjustment, weekly unit maintenance, and so forth. A high-level pseudocode description for the specific SMC method used in our *NARP* tool is shown in Algorithm-1.

| **Algorithm 1: Sequential Monte Carlo Simulation** |
|---|
| 1 **Input**: system data and failure rates of generators and lines |
| 2 **Initialize:** the event list **E** (i.e., a 1D array) |
| 3      insert "event" (i.e., a 2-element tuple (*time*, *event_type*)) *sequentially* for the 1st hour, week, quarter, and year; |
| 4 **While (True)**: |
| 5      **if** *current_list_pos* == **0**: early return; // or throw exception |
| 6      **elif** *current_clock* > time of the last *unhandled* event: |
| 7          early return; // or throw exception |
| 8      **elif** *return_flag* == **0**: normal return; // converged |
| 9      **elif** *current_clock* > FINISN_CLOCK: |
| 10         call function *report*(); // print the final results |
| 11         normal return; // if the maximum clock is reached |
| 12      **else**: |
| 13         fetch the *earliest* unhandled event from the list **E**; |
| 14         call the corresponding handlers based on *event_type* (e.g., call the **OPF-based load shedding** program); |
| 15         set *return_flag* to **1** if reaching convergence criteria; |
| 16         update *current_list_pos* and *current_clock*; |
| 17 **Output**: output files (txt/csv); or failure messages if diverge |

In the above pseudocode, *event_type* is an integer taking value from 1 to 4; *current_clock* and *current_list_pos* are respectively the virtual time clock and the current position pointer (an integer) of the event list.

*2) OPF-based Load Shedding*

The OPF-based load-shedding model of this tool is shown in Eq. (3), where $\Delta P_i^D$ is the decision variable, i.e., the load-shedding amount. $N$, $N_L$, and $E$ are respectively the index sets of all buses, load buses, and branch pairs. The meanings of other symbols are similar to those in a typical OPF formulation (cf. [10] for more details).

$$\begin{aligned}
\min_{V, \text{Re}(S^G), \Delta P_i^D} & \sum_{i \in N_L} \Delta P_i^D \\
\text{s.t.} \quad & P_{ij} = B_{ij}(\theta_i - \theta_j) & \forall (i,j) \in E \\
& P_i^G - P_i^D + \Delta P_i^D = \sum P_{ij} & \forall i \in N, (i,j) \in E \\
& |P_{ij}| \leq P_{ij}^{\max} & \forall (i,j) \in E \\
& P_i^{G,\min} \leq |P_i^G| \leq P_i^{G,\max} & \forall i \in N \\
& \theta_{ij}^{\min} \leq (\theta_i - \theta_j) \leq \theta_{ij}^{\max} & \forall (i,j) \in E
\end{aligned} \quad (3)$$

In this model, the goal is to minimize the load-shedding amount (if necessary), equivalent to minimizing the *power demand not served* in one "hour" step of the simulation.

*3) Unit Preventive Maintenance Schedule*

It is a common practice in real power utilities to schedule preventive maintenance (also called *planned outages*) for units. The goal here is to *levelize* the reserve curve (i.e., "move the capacities" of those *in-maintenance* units onto the annual load profile in proper time intervals) and simultaneously satisfy other constraints. For example, in our tool [11][12]:

- Each unit is up to two planned outages per year
- No outages are allowed in the given *forbidden periods*
- Outage duration is in integer number of weeks

Then, the algorithm in our tool proceeds as follows:

*a) Compute* the product of the *effective capacity* (cf. Eq. (4) and (5)) and outage duration for each unit. Then for each plant, sum these products for all units within it.

$$EC_i = \text{effective capacity of unit } i = C_i - M \log R_i \quad (4)$$

$$R_i = 1 - FOR_i(1 - e^{C_i/M}) \quad (5)$$

where $M$ = slope of the "capacity outage-probability" curve [1]; $FOR_i$ = forced outage rate of unit $i$; $C_i$ = capacity of unit $i$.

*b) Rank* plants based on the above product sums; the larger, the higher. Then, ranking units within each plant based on the product of each unit's rated capacity and maintenance duration; the larger, the higher.

*c) Schedule* units (within each plant) sequentially based on the above priority rank. This scheduling process can minimize the sum of weekly peak load and effective capacity on planned outages while satisfying the power balance constraints.

## III. INPUT AND OUTPUT DESCRIPTIONS

### A. Input Files and Formats

The input data includes several independent sections (or files if stored separately). They can be in either ".txt" or ".csv" format. Each data section or file is called a "data card", introduced one by one below.

- *ZZMC* card: means "miscellaneous data", specifying parameters for the Monte Carlo simulation.

TABLE I. EXAMPLE OF *ZZMC* CARD

| SEED | LS | W1 | W2 | W3 | WHERE | WHEN | KVS | KVT | KVL |
|---|---|---|---|---|---|---|---|---|---|
| 345237 | 1 | 13 | 26 | 39 | 1 | 1 | 1 | 2 | 1 |
| CVT | FIN | STEP | FREQ | MAXE | II | IJ | IR | IN | D | M |
| 0.025 | 9999 | 1 | 1 | 1000 | 0 | 0 | 1 | 5 | 1 | 1 |

An example row is shown in Table I, where meanings of certain key parameters are (cf. [12] for remainders):

**SEED** – the random number seed (positive integer);

**LS** – an integer indicator of "loss-sharing mode" (if 0) or "non-loss sharing mode" (if 1) (cf. [10]-[12]).

**W1**, **W2**, **W3** – the integer indices of the ending weeks for the first three seasons (quarters).

**WHERE** – if 0, convergence is judged based on statistics of one fixed area (specified by **KVL**). If 1, convergence is judged based on the pool statistics.

**CVT** – the threshold for the converge criteria $\beta$ in Eq. (2).

**FIN** – the maximum simulation runs ("years")

- *ZZLD* card: means "area data", specifying parameters for each area of the interconnected system.

TABLE II. EXAMPLE ROW OF *ZZLD* CARD

| SN | AREA NAME | PEAK (MW) | LFU (%) | OUTAGE WINDOW | | FORBIDDEN PERIOD | | SUM OF FLOWS CONSTRAINT |
|---|---|---|---|---|---|---|---|---|
| | | | | BEG WK | END WK | BEG WK | END WK | |
| 1 | 'A1' | 3000 | 0 | 1 | 52 | 31 | 32 | 30000 |

An example row is shown in Table II, where:

**SN** – the serial number identifies each entry uniquely and specifies the sequence of data entries.

**Area Name** – the name of the area. It can have up to four letters and must be enclosed with quotes.

**Peak** – annual peak in the area (MW).

**LFU** – load forecast uncertainty (%), one standard deviation expressed as a percentage of the mean value.

**Outage Window** – the weeks during which planned maintenance of generator units can be performed, defined via the integer indices of the beginning week (**BEG WK**) and the ending week (**END WK**).

**Forbidden Period** – a part of the outage window during which planned maintenance is not allowed, also defined by two integer indices, **BEG WK** and **END WK**.

**Sum of Flows Constraint** – the algebraic sum of flows at this area cannot exceed this value.

- *ZZUD* card: means "generator unit data", specifying generator unit parameters.

TABLE III. EXAMPLE ROW OF *ZZUD* CARD

| SN | UN NAME | LOC | CAP1 | CAP2 | CAP3 | CAP4 | DFOR |
|---|---|---|---|---|---|---|---|
| 1 | 'A10101' | 'A1' | 12 | 12 | 12 | 12 | 0 |
| FOR | DER | P/A | B1 | D1 | B2 | D2 | |
| 0.02 | 0 | 0 | 0 | 0 | 0 | 0 | |

An example row is shown in Table III, where:

**UN Name** – the alphanumeric name of each generator unit. The first four letters are the plant number. The next two digits identify each unit within that plant.

**LOC** – the area of location, up to four letters.

**CAP**$i$ – the generator unit capacity (MW) in the $i$th season (cf. *ZZLD* card)

**DFOR** – the forced outage rate corresponding to the specific unit's derated (i.e., partial) capacity loss.

**FOR** – the forced outage rate corresponding to the full capacity loss of the specific unit.

**DER** – the percent (%) of derated capacity due to partial capacity loss of the specific unit.

**P/A** – predetermined (if 1) or automatic scheduling (if 0).

**B1**, **D1** – the beginning week and duration of the first outage. In the automatic scheduling mode, B1 is ignored.

**B2**, **D2** – the beginning week and duration of the second outage. In the automatic scheduling mode, B2 is ignored.

- *ZZFC* card (optional): means "firm contracts data", specifying parameters for the firm power (MW) interchanges between pairs of areas.

TABLE IV. EXAMPLE ROW OF *ZZFC* CARD

| SN | FROM AREA | TO AREA | BEG DAY | END DAY | MW |
|---|---|---|---|---|---|
| 1 | 'A1' | 'A2' | 14 | 15 | 630 |

An example row is shown in Table IV, where the meanings of the first three columns are similar to that of the above cards. The last three columns indicate how much (contracted) power must be exchanged between the specified two areas on each day of a pre-contracted time window.

- *ZZOD* card (optional): means "unit ownership data", specifying joint ownerships of a unit across areas.

TABLE V. EXAMPLE ROW OF *ZZOD* CARD

| SN | UNIT NAME | PERCENT OWNED BY AREA |
|---|---|---|
| 1 | 'A10101' | 10,20,30,40,0,0,0,0,0,0,0,0,0,0 |
| 2 | 'A10102' | 0,0,10,20,30,40,0,0,0,0,0,0,0,0 |

An example row is shown in Table V, where the meanings of the first two columns are similar to that of the above cards.

The last column indicates the capacity shares of a unit owned by each area. The sum should be equal to 100.

- *ZZTD* card means "transmission line data", specifying parameters for interarea tie-lines.

TABLE VI. EXAMPLE ROW OF *ZZTD* CARD

| SN | LINE NUMBER | FROM AREA | TO AREA | META DATA |
|---|---|---|---|---|
| 1 | 1 | 'A1' | 'A2' | "-120, 300, 300, 0.9216, -60, 150, 150, 0.0768, 0, 0, 0, 0.0016, -80, 150, 150, 0.0000, -40, 100, 100, 0.0000, -20, 50, 50, 0.0000" |

An example row is shown in Table VI, where the meanings of the first four columns are similar to that of the above cards. Each row represents a specific line, and the "Line Number" is *not* necessarily the same as the "SN". The "Meta Data" column is a long string. Its meaning is as follows:

"**ADM**, **CAP**, **CAPR**, **PROBL**"

where, for each line, it has up to six possible states. For each state, the line *can* have different values of admittance ("**ADM**", in per unit), power flow capacity ("**CAP**", in MW), reverse power flow capacity ("**CAPR**", in MW), and probability (**PROBL**) of that state.

### B. Output Files and Formats

The output file can be either one "txt" file including all the tables of results or several "csv" files. Those result tables can be categorized as follows:

#### 1) System Re-description

This part reiterates both system meta-information and settings of the Monte Carlo simulation. For example, the peak load in each area; each unit's string name, location, force-outage rate; the maximum "years" to be simulated (if no convergence is reached); the frequency by which the Monte Carlo statistics are collected (hourly or daily), and so on.

#### 2) Weekly Schedule of Unit Maintenance

For each area, this part will show three tables: 1) "weekly peak load before maintenance schedule", a 4-by-13 table listing the given annual load profile; 2) "planned maintenance schedule for area-*i*". For example, suppose two units (named "A10101" and "A13201") in Area-1 and one unit (named "A23201) in Area-2 *must* be arranged for a two-week maintenance (specified by the *ZZUD* card). Then our tool can find a proper schedule as shown in Fig. 3 and 4 for Area-1 and 2 (where "AA" means maintenance is automatically scheduled for a specific week). Also, the adjusted annual load profile (using the *week index* as the *x*-axis) is shown in Fig. 5.

Figure 3. The obtained unit maintenance schedule for two units in Area-1

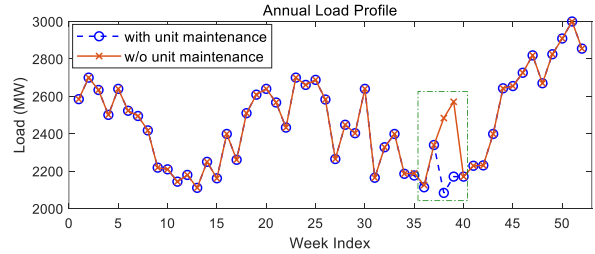

Figure 4. The obtained unit maintenance schedule for a unit in Area-2

Figure 5. Comparison of load profiles: with and without unit maintenance

#### 3) Statistics of Reliability Indices

In this part, area-wise tables will be presented, where EUE (expected unserved energy, MWh), XLOL (loss of load, MW), LOLE (loss of load expectation, days/yr), and HLOLE (hourly LOLE, hrs/yr) are listed separately as the statistics. Moreover, the above results are further segmented into three sub-categories: generation inadequacy-caused (marked as "GC"), transmission inadequacy-caused (marked as "TC"), and their combination (marked as "GT"). Finally, another table will present the overall statistics for the entire system (called "pool statistics"). Refer to Section IV for examples.

#### 4) Probability Distributions of Reliability Indices

In this part, area-wise probability distributions are listed in tables under three categories: "Daily peak LOLEs per year", "HLOLEs per year", and "Annual unserved energy (MWh)". Each table has 20 rows of numeric values, representing 20 bins used in characterizing the probability distributions. Refer to Section IV for examples.

## IV. EXAMPLE CASE STUDY

In this section, a five-area system is adopted to present an example case study using our software. The data can be downloaded from [7]. Its settings are:

- *Each area is an equivalent of a revised RTS-79 system*
- *120 generator units; 32 units per each area*
- *5 interarea tie-lines*
- *Peak load of each area: 3000 MW*
- *Sum of flows constraint in each area: 30000 MW*

The hyperparameters and settings for the Monte Carlo simulation are:

- *Convergence criteria*: $\beta = 0.025$
- *Max simulation runs*: 9999
- *Hourly statistics are collected during the simulation*
- *HLOLE is used for convergence*
- *The area used for convergence check: the 1st area*.

The program converged in about 6 minutes (at the 3576-th "year") for this system on a laptop with a 4GHz CPU and 32GB RAM, which is much faster than similar experiments implemented in MATLAB. Regarding the output results, the unit maintenance results have been displayed in the previous section. Therefore, the following part will mainly present the results for 1) reliability statistics and 2) reliability probability distributions.

## A. Results of Reliability Statistics

The screenshots of reliability statistics results are shown in Fig. 6. From the results of this specific case, it can be observed that the generator failures contribute more than line failures in terms of the final LOLE, HLOLE, and EUE.

```
              |           FINAL RESULTS AFTER    3576 REPLICATIONS
              |
AREA FORECAST |        HOURLY STATISTICS              PEAK STATISTICS       REMARKS
 NO    NO     | HLOLE      XLOL       EUE         LOLE        XLOL
              |(HRS/YR)    (MW)       (MWH)      (DAYS/YR)    (MW)
  1    AV       0.10      132.99      14.          0.04      130.21         GC
  1    AV       0.35      103.12      36.          0.07      111.29         TC
  1    AV       0.45      110.04      50.          0.10      117.97         GT

  2    AV       0.11      134.66      15.          0.03      124.76         GC
  2    AV       0.33      102.63      34.          0.06      108.77         TC
  2    AV       0.44      110.61      48.          0.09      114.45         GT

  3    AV       0.11      147.01      16.          0.04      158.24         GC
  3    AV       0.39      103.38      41.          0.08      110.23         TC
  3    AV       0.50      112.82      57.          0.12      126.31         GT

  4    AV       0.10      149.40      16.          0.04      141.96         GC
  4    AV       0.34       98.52      33.          0.06      101.61         TC
  4    AV       0.44      110.55      49.          0.10      116.22         GT

  5    AV       0.10      127.32      13.          0.03      138.64         GC
  5    AV       0.31       96.19      30.          0.06      106.66         TC
  5    AV       0.41      104.03      43.          0.09      118.71         GT

                              POOL STATISTICS
       AV       0.245     300.16      74.          0.080     309.59         GC
       AV       1.634     106.11     173.          0.305     114.52         TC
       AV       1.879     131.41     247.          0.385     155.18         GT
```

Figure 6. Screenshots of obtained reliability statistics

## B. Results of Reliability Probability Distribution

Due to the page limit, probability (density) distributions of LOLE, HLOLE, and EUE for Area-1 and Area-2 are selected to show in Fig. 7. Histograms of those reliability indices are also depicted in Fig. 8 for better visualization.

Figure 7. Probability distributions of (H)LOLE, EUE for Area-1 and Area-2

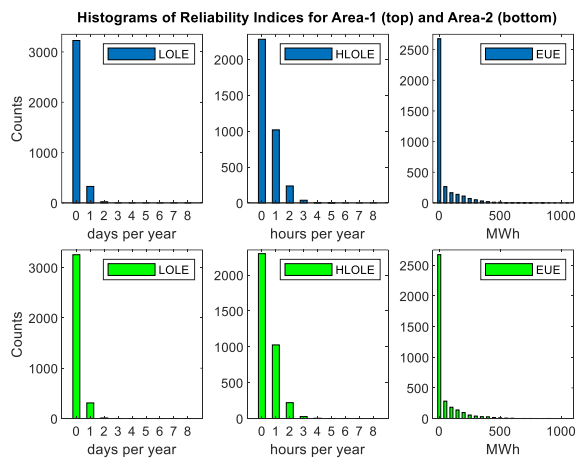

Figure 8. Histograms of (H)LOLE, EUE for Area-1 and Area-2

## V. CONCLUSION AND FUTURE WORK

A free industry-grade education tool for bulk-power-system reliability assessment has been developed. Sequential Monte Carlo simulation is utilized, and advanced features due to real-world needs can be considered during the simulation. Due to the Fortran implementation of its internal algorithms, a rapid computation speed is achieved even under small convergence thresholds. Results of area-wise reliability indices can be displayed in tables and plots. The next steps are: 1) implementing other advanced algorithms (e.g., variance reduction techniques) and 2) enhancing the GUI.